\documentclass[twocolumn,showpacs,aps,amsmath,amssymb,prb,preprintnumbers]{revtex4}
\newcommand{\BE}{\begin{equation}}
\newcommand{\EE}{\end{equation}}
\newcommand{\BA}{\begin{eqnarray}}
\newcommand{\EA}{\end{eqnarray}}
\usepackage{graphicx}
\usepackage{dcolumn}
\usepackage{bm}
\input epsf.tex

\begin{document}

\title{Guiding optical flows by photonic crystal slabs made of dielectric cylinders}

\author{Liang-Shan Chen$^1$} \author{Chao-Hsien Kuo$^2$}\author{Zhen
Ye$^2$}\email{zhen@phy.ncu.edu.tw} \affiliation{$^1$Department of
Physics, Fudan University, Shanghai, China, and $^2$Wave Phenomena
Laboratory, Department of Physics, National Central University,
Chungli, Taiwan}

\date{\today}

\begin{abstract}

We investigate the electromagnetic propagation in two-dimensional
photonic crystals, formed by parallel dielectric cylinders
embedded a uniform medium. The frequency band structure is
computed using the standard plane-wave expansion method, while the
propagation and scattering of the electromagnetic waves are
calculated by the multiple scattering theory. It is shown that
within partial bandgaps, the waves tend to bend away from the
forbidden directions. Such a property may render novel
applications in manipulating optical flows. In addition, the
relevance with the imaging by flat photonic crystal slabs will
also be discussed.

\end{abstract}

\pacs{78.20.Ci, 42.30.Wb, 73.20.Mf, 78.66.Bz} \maketitle

\begin{small}

\section{Introduction}

When propagating through periodically structured media such as
photonic crystals (PCs), optical waves will be modulated with the
periodicity. As a result, the dispersion of waves will no longer
behave as in a free space, and so called frequency band structures
appear. Under certain conditions, waves may be prohibited from
propagation in certain or all directions, corresponding to partial
and complete bandgaps respectively. The photonic crystals
revealing bandgaps are called bandgap materials.

Photonic crystals and band gap materials have a broad spectrum of
applications, ranging from computing to digital communication and
from laser cavities to optical transistors\cite{Book}. The
possibilities are unlimited. In fact, applications have well gone
beyond expectation, and are so far reaching that a fruitful new
field called photonic crystals has come into existence. Most
updated information about the research of photonic crystals and
related materials can be found in the comprehensive webpage
\cite{web}.

So far, most applications are associated with the properties of
the complete bandgaps of PCs. On one hand, the bandgaps confine
optical propagation within certain frequency regimes. On the
other, when encountering the complete bands, optical waves can be
guided into desired directions. For example, one of the main
applications of PCs is to control optical flows, so that they can
be used for such as telecommunications. A comprehensive survey of
phonic crystal research can be referred to
Refs.~\cite{Book,web,Mono1,Mono2,Mono3}. To our knowledge,
however, there have been very few attempts in the literature to
explore possible usage of partial bandgaps. In this paper, we wish
to discuss a previously undiscussed phenomenon associated with
partial bandgaps, that is, deflection of optical waves. That is,
the partial bandgap can collimate wave propagation into certain
directions. This property may allow for novel applications in
manipulating optical flows.

The paper is organized as follows. The systems and the theory will
be outlined in the next section. The results and discussion will
be presented in Section III, followed by a short summary.

\section{The systems and formulation}

The systems considered here are two dimensional photonic crystals
made of arrays of parallel dielectric cylinders placed in a
uniform medium, which we assume to be air. Such systems are common
in both theoretical simulations or experimental measurements of
two dimensional PCs\cite{Book,web}. For brevity, we only consider
the E-polarized waves (TM mode), that is, the electric field is
kept parallel to the cylinders. The following parameters are used
in the simulation. (1) The dielectric constant of the cylinders is
14, and the cylinders are arranged to form a square lattice. (2)
The lattice constant is $a$ and the radius of the cylinders is
0.3$a$; in the computation, all lengths are scaled by the lattice
constant. (3) The unit for the angular frequency is $2\pi c/a$.
After scaling, the systems become dimensionless; thus the features
discussed here would be applicable to a wider range of situations.

While the frequency band structure in the systems can be
calculated by the plane-wave expansion method\cite{Book}, the
propagation and scattering of electromagnetic (EM) waves in such
systems can be studied by the standard multiple scattering theory.
The theory originated from the self-consistent idea first
discussed by Foldy\cite{Foldy}, and then made maturity through the
significant efforts by Lax\cite{Lax}, Waterman et
al.\cite{Waterman}, and particularly by Twersky\cite{Twersky}.

The essence of the theory is summarized as follows. In response to
the incident wave from the source and the scattered waves from
other scatterers, each scatter will scatter waves repeatedly, and
the scatterered waves can be expressed in terms of a modal series
of partial waves. When this scattered wave serves as an incident
wave to other scatterers, a set of coupled equations can be
formulated and computed rigorously. The total wave at any spatial
point is the summation of the direct wave from the source and the
scattered waves from all scatterers. The intensity of the wave is
represented by the square of the wave field.

For the reader's convenience we present briefly the general
multiple scattering theory. Consider that $N$ straight cylinders
of radius $a^i$ located at $\vec{r_i}$ with $i = 1, 2,...,N$ to
form an array. A line source transmitting monochromatic waves is
placed at $\vec r_s$. Here we take the standard approach with
regard to the source. That is, the transmission from the source is
calculated from the multiple scattering theory, and assume that
the source is not affected by the surroundings. If some other
sources such as a line of atoms are used, the reaction between the
source and the backscattered waves should be taken into account.

The scattered wave from each cylinder is a response to the total
incident wave composed of the direct wave from the source and the
multiply scattered waves from other cylinders. The final wave
reaching a receiver located at $\vec r_r$ is the sum of direct
wave from the source and the scattered waves from all the
cylinders.

The scattered wave from the $j$-th cylinder can be written as
\begin{equation}
p_s(\vec{r}, \vec{r_j}) = \sum_{n=-\infty}^\infty i \pi A_n^j
H_n^{(1)}(k|\vec{r} - \vec{r_j}|) e^{in\phi_{\vec{r} -
\vec{r_j}}},
\end{equation}
where $k$ is the wavenumber in the medium, $H_n^{(1)}$ is the
$n$-th order Hankel function of first kind, and $\phi_{\vec{r} -
\vec{r_j}}$ is the azimuthal angle of the vector $\vec{r} -
\vec{r_j}$ relative to the positive $x$ axis. The total incident
wave around the $i$-th cylinder $(i=1,2,...,N; i \ne j)$ is the
summation of the direct incident wave from the source and the
scattered waves from all other scatterers, can be expressed as
\begin{equation}
p_{in}^i(\vec r) = \sum_{n=-\infty}^\infty B_n^i J_n(k|\vec r -
\vec{r_i}|) e^{in\phi_{\vec r - \vec{r_i}}}.
\end{equation}
In this paper, $p$ stands for the electrical field in the TM mode
and the magnetic field in the TE mode.

The coefficients $A_n^i$ and $B_n^i$ can be solved by expressing
the scattered wave $p_s(\vec r, \vec{r_j})$, for each $j \ne i$,
in terms of the modes with respect to the $i$-th scatterer by the
addition theorem for Bessel function. Then the usual boundary
conditions are matched at the surface of each scattering cylinder.
This leads to
\begin{equation}
B_n^i = S_n^i + \sum_{j=1, j \ne i}^N C_n^{j,i}, \label{eq:6}
\end{equation}
\noindent with
\begin{equation}
S_n^i = i \pi H_{-n}^{(1)}(k|\vec{r_i}|)e^{-i n \phi_{\vec r_i}},
\end{equation}
and
\begin{equation}
C_n^{j,i} = \sum_{l=-\infty}^\infty i \pi A_l^j
          H_{l-n}^{(1)}(k|\vec{r_i} - \vec{r_j}|)
          e^{i(l-n)\phi_{\vec{r_i} - \vec{r_j}}},
\end{equation}
and
\begin{equation}
B_n^i = i \pi \tau_n^i A_n^i,
\end{equation}
where $\tau_n^i$ are the transfer matrices relating the properties
of the scatterers and the surrounding medium and are given as \BE
\tau_n^i = \frac{H_n^{(1)}(ka^i)J_n^\prime(ka^i/h^i) - g^i h^i
{H_n^{(1)}}^\prime (ka^i)J_n(ka^i/h^i)}{g^ih^iJ_n^\prime({ka^i})
J_n(ka^i/h^i) - J_n(ka^i)J_n^\prime(ka^i/h^i)}, \EE where
$$
 h^i = \frac{1}{\sqrt{\epsilon^i}}, \ \ \ \mbox{and} \ \ \
g^i = \left\{\begin{array}{ll} \epsilon^i & \mbox{for TE waves}\\
1 & \mbox{for TM waves} \end{array}\right.,
$$
in which $\epsilon^i$ is the dielectric constant ratio between the
$i$-th scatterer and the surrounding medium.

The coefficients $A_n^i$ and $B_n^j$ can then be inverted from
Eq.~(\ref{eq:6}). Once the coefficients $A_n^i$ are determined,
the transmitted wave at any spatial point is given by
\begin{equation}
p(\vec r) = p_0(\vec r) + \sum_{i=1}^N \sum_{n=-\infty}^\infty
            i \pi A_n^i H_n^{(1)}(k|\vec r - \vec{r_i}|)
            e^{in\phi_{\vec r - \vec{r_i}}},
\end{equation}
where $p_0$ is the field when no scatterers are present. The
transmitted intensity field is defined as $|p|^2$.

\section{Results and discussion}

The frequency band structure is plotted in Fig.~1, and the
qualitative features are similar to that obtained for a square
array of alumina rods in air. A complete band gap is shown between
frequencies of 0.22 and 0.28. Just below the complete gap, there
is a regime, sandwiched by two horizontal lines, of partial band
gap in which waves are not allowed to travel along the $\Gamma X$
or [10] direction. We will consider waves whose frequency is
within this partial bandgap. In particular, we choose the
frequency to be 0.192.

First we consider the propagation of EM waves through two
rectangular slabs of arrays of dielectric cylinders. Figure~2
shows the images of the fields. The left panel shows the real
parts of the fields $E_z$, while the right panel presents the
images of the intensity fields $|E_z|^2$. In (a1) and (a2), the
slab measures 14x45, and the slab is oriented such that the [11]
direction, i.~e. the $\Gamma M$ direction, is along the horizontal
level. The size of the slab in (b1) and (b2) is 10x45, and the
[11] direction is titled upwards and makes an angle of 22.5 degree
with respect to the horizontal direction. The frequency is chosen
at 0.192. A transmitting point source is placed at 2 lattice
constant away from the left side of the slabs. The detailed
geometrical information can be referred to in Fig.~2.

A few observations can be drawn from Fig.~2. First, there is a
focused image across the slab in (a1) and (a2). Earlier, this
focused image was attributed to the effect of negative
refraction\cite{Ye1}, inferred from the group velocity
calculation. If this conjecture were valid, another focused image
would be expected inside the slab as well. Our result does not
support this conjecture. As seen from (a1) and (a2), there is no
focused image inside the slab. Rather, the waves are mostly
confined in a tunnel and travel to the other side of the slab,
then release to the free space. This is understandable, because
the forbidden direction in (a1) and (a2) is along $\Gamma X$,
which makes an angle of 45 degree from the $\Gamma M$ direction
that lies horizontally. The passing band in the $\Gamma M$
direction thus acts as a transportation carrier that moves the
source to the other side of the slab. The waves on the right hand
side of the slab look as if they were radiated by an image that
has been transported across the slab within a narrow guide.
Second, the waves tend to bend towards the $\Gamma M$ direction,
as evidenced by Fig.~2 (b1) and (b2). Third, the decay of the
transported intensity along the travelling path is not obvious, an
indication of efficient guided propagation.

The results in Fig.~2 are promising. They show that in the
presence of partial bandgaps and when incident upon a slab of
photonic crystals, waves tend to bend toward directions which are
mostly away from forbidden directions. This would indicate that
partial bandgaps may be considered as a candidate for guiding wave
flows. To verify this conjecture, we have further explored the
guiding phenomenon associated with partial bandgaps.

In Fig.~3, we show the EM wave propagation through stacks of
photonic crystal slabs. Two setups are considered. In (a1) and
(a2), two slabs of dielectric cylinders are stacked together. The
first (left) slab is oriented such that the [11] direction is
horizontal, while the second (right) slab is arranged to make the
[11] direction tilted upward, making 22.5 degree with respect to
the horizontal direction. The two slabs measure as 9x44 and 14x44
respectively. In (b1) and (b2), two slabs are adjacently attached.
The [11] direction is tilted upward by 10 degree for the first
(left) slab, while it is along the horizontal direction in the
second (right) slab. The sizes of the two slabs are 8x40 and 10x40
respectively. In both situations (a) and (b), the point source is
placed at a distance of 1.5 away the left side of the stacks. The
purpose here is to show how the light would travel when two
adjacent slabs have different orientations.

Here it is clearly shown that the EM waves indeed always tend to
travel along the [11] direction. In the case of (b1) and (b2), an
image has been formed on the right hand side. Compared to the
source, the image is uplifted by a distance about $8\tan(\pi/18)$.
In the cases considered here, the first slabs (left) serve as a
collimating device, and then the collimated waves will be guided
by subsequent photonic crystal slabs. This consideration can be
extended to multiple consecutive slabs so that the wave flows can
be guided into desired orientations, making possible alternative
ways of controlling optical flows.

In Fig.~4, we consider two other situations of stacked photonic
crystal slabs. The geometrical parameters are indicated in the
figure. Again, the waves tend to move along the [11] direction.
Here the amphoteric diffraction is observed. It draws analogy with
the amphoteric refraction observed when waves propagate from an
isotropic to an anisotropic medium\cite{Yau}.

The results from Figs.~2, 3 and 4 clearly indicate that the
partial bandgaps can be indeed used as a guiding channel for
optical flows. It can be also inferred that the guided transport
is efficient. We have carried out further simulations against
variations of frequencies, filling factors, and dielectric
constants, the results are quantitatively the same for waves
within partial bandgaps. The observation presented here has also
been confirmed by FDTD simulations. The controlled wave transport
due to partial bandgaps of PCs should be interpreted in terms of
diffraction or scattering rather than refraction; in fact, no
refraction index can be determined for the phenomenon.

An immediate question may thus arise: Why the waves of frequencies
within the partial bandgap tend to bend to particular directions?
To answer this question, we have examined the properties of the
energy flow of the eigenmodes which correspond to the frequency
bands. While details will be published elsewhere, here we only
outline our thoughts. The usual approach mainly relies on the
curvatures of frequency bands to infer the energy flow. As
documented in Ref.~\cite{Yeh}, an energy velocity is defined as
$\vec{v}_e = \frac{\frac{1}{V}\int \vec{J}_{\vec{K}}
d^3{r}}{\frac{1}{V}\int U_{\vec{K}} d^3{r}},$ where
$\vec{J}_{\vec{K}}$ and $U_{\vec{K}}$ are the energy flux and
energy density of the eigenmodes, and the integration is performed
in a unit cell. It can be shown that thus defined energy velocity
equals the group velocity obtained as $\vec{v}_g =
\nabla_{\vec{K}}\omega(\vec{K}).$ Therefore it is common to
calculate the group velocity to infer the energy velocity and
subsequently the energy flows or refraction of waves. A few
questions, however, may arise with regard to this approach. First,
when the variation in the Bloch vector, i.~e. $\delta \vec{K}$, is
small, the changes in $\omega$, $\vec{E}_{\vec{K}}$ and
$\vec{H}_{\vec{K}}$ should also be small. Second, the variation
operation should be exchangeable with the partial differential
operations. When these two conditions fail, the energy velocity
will become illy defined. Third, even if the two conditions hold,
whether the net current flow through a unit cell really follows
the direction of $\vec{v}_e$ remains unclear. We note here that
the average flux through a surface may be defined as
$\langle\vec{J}\rangle = \frac{\hat{n}}{S}\int d\vec{S}\cdot
\vec{J}$, where $\hat{n}$ is the unit normal vector of the surface
$S$. Clearly, the volume averaged current within a unit cell does
not necessarily correspond to the actual current flow. We will
publish verifications elsewhere.

To avoid possible ambiguities, here we consider the energy flow
based upon its genuine definition. One advantage of this approach
is that we are also able to examine the local properties of energy
flows. By Bloch's theorem, the eigenmodes corresponding to the
frequency bands of PCs can be expressed as $E_{\vec{K}}(\vec{r}) =
e^{i\vec{K}\cdot\vec{r}}u_{\vec{K}}(\vec{r})$, where $\vec{K}$ is
the Bloch vector, as the wave vector, and $u_{\vec{K}}(\vec{r})$
is a periodic function with the periodicity of the lattice. When
expressing $E_{\vec{K}}(\vec{r})$ as
$|E_{\vec{K}}(\vec{r})|e^{i\theta_{\vec{K}}(\vec{r})}$, the
corresponding energy flow is derived as
$\vec{J}_{\vec{K}}(\vec{r}) \propto
|E_{\vec{K}}(\vec{r})|^2\nabla\theta_{\vec{K}}(\vec{r})$; clearly
$\theta_{\vec{K}}$ combines the phase from the term
$e^{i\vec{K}\cdot\vec{r}}$ and the phase from the function
$u_{\vec{K}}(\vec{r})$. To explore the characteristics of the
partial bandgap, we have computed the eigen-field
$E_{\vec{K}}(\vec{r})$ and also the energy flow of the eigenmodes.
The results are shown in Fig.~5. Fig.~5(a) shows that the energy
eventually tends to flow into the direction of $\Gamma M$, i.~e.
the [11] direction, while the Bloch vector points to an angle of
22.5$^o$ that lies exactly between $\Gamma X$ and $\Gamma M$.

Although the above features are only investigated for the first
partial bandgap in this paper, we have found that they are also
valid for other partial bandgaps. For example, we have also
considered the second partial bandgap which is located between
0.283 and 0.325. All above features remain quantitatively valid.
Within this second gap, however, the waves are collimated to
travel along the [10] direction rather than the [11] direction. In
addition, we have also carried out simulations for various slab
sizes, all the features are the same, thus excluding the
boundaries as the possible cause.

\section{Summary}

We have considered EM wave propagation through slabs of photonic
crystals which are made of arrays of dielectric cylinders.
Properties of partial bandgaps are investigated. It was shown that
the partial bandgaps may act as a guiding channel for wave
propagation inside the photonic crystals. Such a feature may lead
to novel applications in manipulating optical flows.

\section*{Acknowledgments}

The work received support from National Science Council, National
Central University, and Fudan University.


\section*{Figure Captions}

\begin{description}

\item[Fig. 1] The band structure of a square lattice of dielectric
cylinders. The lattice constant is $a$ and the radius of the
cylinders is $0.3a$. $\Gamma M$ and $\Gamma X$ denote the [11] and
[10] directions respectively. A partial gap is between the two
horizontal lines.

\item[Fig. 2] Imaging of the transmitted fields across two slabs
of dielectric cylinders. The black circles denote the cylinders
(for clarity, not all cylinders are plotted).

\item[Fig. 3] Imaging of the intensity fields across
two-consecutive slabs of arrays of dielectric cylinders in two
arrangements. The left and right panels respectively plot the real
part of the field and the intensity.

\item[Fig. 4] Imaging of the intensity fields across
two-consecutive slabs of arrays of dielectric cylinders in two
arrangements. The left and right panels respectively plot the real
part of the field and the intensity. Here is show the amphoteric
diffraction at the interfaces between two adjacent slabs: (a)
positive and (b) negative. In both (a) and (b), the adjacent slabs
measure as 10x50 and 12x50.

\item[Fig. 5] Left panel: the field pattern of eigenmodes. Right
panel: the energy flow of the eigenmodes. The eigenmodes along two
directions are considered: (a) $\vec{K} = (0.9\pi/a, 0.37\pi/a)$,
i.~e. along an angle of 22.5$^o$ exactly between $\Gamma X$ and
$\Gamma M$ directions; the corresponding frequency is 0.185; (b)
$\vec{K} = (0.7\pi/a, 0.7\pi/a)$, i.~e. along $\Gamma M$; the
corresponding frequency is 0.192. The direction of the Bloch
vectors are denoted by the blue arrows, while the red arrows
denote the local energy flow including the direction and the
magnitude. The circles refer to the cylinders. Both frequencies in
(a) and (b) lie within the partial gap. Due to the periodicity, we
only plot the energy flow within one unit cell. Note that although
the features shown by (a) also hold for other Bloch vectors for
which the corresponding frequencies lie within the partial gap
regime, we only plot here for the case of 22.5$^o$.

\end{description}

\end{small}

\end{document}